\documentstyle[preprint,prl,aps,epsf]{revtex}
\begin{document}
\title{Nonresonant corrections to the 1s-2s two-photon resonance\\
for the hydrogen atom.}

\author{ L. Labzowsky$^{1,2}$, D. Soloviev$^{1}$, G. Plunien$^{3}$,
G. Soff$^{3}$}
\address{ $^1$St. Petersburg State University, 198504, Petrodvorets, 
St. Petersburg, Russia} 
\address{ $^2$Petersburg Nuclear Physics Institute, 188350, Gatchina, 
St. Petersburg, Russia}
\address{ $^3$Technische Universit$\ddot{a}$t Dresden, Mommsenstr. 13,
 D-01062, Dresden, Germany}
\maketitle

\begin{abstract}
The nonresonant (NR) corrections are estimated for the most
accurately measured two-photon transition 1s-2s in the hydrogen
atom. These corrections depend on the measurement process and set
a limit for the accuracy of atomic frequency measurements.
With the measurement process adopted in the modern experiments
the NR contribution for 1s-2s transition energy can reach $10^{-3}$
Hz while the experimental inaccuracy is quoted to be $\pm$ 46 Hz.
\end{abstract}

\bigskip
Nonresonant (NR) corrections have been first introduced in
Ref. \cite{Low} where the modern QED theory of the natural line profile
in atomic physics has been formulated. The NR corrections indicate the
limit up to which the concept of the energy of an excited atomic
state has a physical meaning - that is the resonance
approximation. In the resonance approximation the line profile is
described by the two parameters: energy $E$ and width $\Gamma$ .
Beyond this approximation the specific role of $E$ and $\Gamma$ should
be replaced by the complete evaluation of the line profile for the
particular process. If the distortion of the Lorentz profile is
small one can still consider the NR correction as an additional
energy shift. Unlike all other energy corrections, this correction
depends on the particular process which has been employed for the
measurement of the energy difference. 
Quite independent of the accuracy
of theoretical calculations of the "traditional" energy corrections
which can be much poorer than the experimental accuracy, NR
corrections define the principal limit for the latter. 
One can state that nonresonant 
corrections set the limit for the accuracy of all the atomic
frequency standards.

Nonresonant corrections have been evaluated for H-like ions of phosphorus
(Z=15) and uranium (Z=92) in Refs. \cite{LKG94,LGaL97}. While for
uranium the NR correction turned out to be negligible, its value
was comparable with the experimental error bars in case of
phosphorus. Recently the NR correction has been evaluated for the
Lyman-$\alpha$ 1s-2p transition in hydrogen \cite{LSPS01}. In
\cite{LSPS01} the process of the resonance photon scattering was
considered as a standard procedure for the determination of the
energy levels. According to \cite{Low} the parametric estimate of
the NR correction to the total cross-section is (in relativistic
units):
\begin{equation}
\delta_{1}=c_{1}m\alpha^{2}(\alpha Z)^{6}
\end{equation}
Here $\alpha$ denotes the fine structure constant, $m$ is the electron
mass, $Z$ is the nuclear charge number and $c_{1}$ abbreviates a numerical
factor. Evaluations in \cite{LSPS01} yielded 
$c_{1} = -1.31*10^{-3}$
and $\delta_{1} = -1.30$ Hz. In Ref. \cite{LSPS01} a factor $4/9$ was
missing in the expression for $c_{1}$. Moreover, there exists
another NR correction to the total cross section of the  same
process, connected with the contribution of the near-resonant
$2p_{3/2}$ state (see below):
\begin{equation}
\delta_{2}=c_{2}m\alpha^{4}(\alpha Z)^{4}
\end{equation}
For the hydrogen atom ($Z=1$) this correction is parametrically of
the same order, but numerically it is larger than (1): 
$c_{2}=4.94\, 10^{-3}$,
$\delta_{2}=4.89$ Hz. Finally, the "asymmetry" correction, which is of the
same order, arises from the resonant term when replacing the width
$\Gamma (\omega_{0})$, where $\omega_{0}$ is the resonant
frequency, by $\Gamma (\omega)$. In the case of the Lyman-alpha
transition $\Gamma
(\omega_{0})=\left(\frac{2}{3}\right)^{8}\alpha^{3}$ a.u. should
be replaced by $\Gamma (\omega)=(2^{11}/3^{9})\alpha^{3}\omega$ a.u.
This result corresponds to the "velocity" form of the transition
amplitude. The "length" form that follows from the "velocity" form
after the gauge transformation would yield $\Gamma
(\omega)=(2^{17}/3^{11})\alpha^{3}\omega^{3}$ a.u. The
"asymmetry" shift is different for the "velocity" and the "length"
forms. However, we should remember that the "velocity" form
follows originally from the QED description and that the gauge
transformation in the simple form mentioned above does exist only
at the resonance frequency. The additional "asymmetry" shift which
follows from the "velocity" form of $\Gamma (\omega)$ is given by 
$\delta_{3}=c_{3}m\alpha^{2}(\alpha Z)^{6}$ with 
$c_{3}=\left(\frac{2}{3}\right)^{17}=1.015*10^{-3}$   
 yielding $\delta_{3}=1.007$ Hz. 
The $\omega$-dependent terms in the Lorentz denominator lead to smaller
NR contributions \cite{Low}. Still the accuracy for the
Lyman-$\alpha$ frequency measurements is much poorer: about $6$ MHz
\cite{Eihema01}. Modern experimental techniques employed in 
Lamb-shift measurements are based on two-photon resonances, in
particular for the transition 1s-2s \cite{Hunder99,Niering00}. In
the present paper we provide an estimate for the NR correction to the
transition frequency in this case.

The magnitude of the NR corrections depends strongly on the
conditions of the experiment. It is important whether total or
differential cross sections are measured since some NR corrections
vanish after the angular integration. 
In \cite{LSPS01} the following expression for the NR correction to the 
frequency of the one-photon transition, which applies 
in case when the total cross section scattering is measured,  
has been derived:

\begin{equation}
\delta_{1}=-\frac{1}{4} \frac{\Gamma_{A'}^{2}}{\Gamma_{A'A}
 \Gamma_{BA'}}Re\left[\sum_{n \neq A'}\frac{\Gamma_{AA;A'n} \Gamma_{BB;nA'}}{E_{n}-E_{A'}}+\sum_{n'}
  \frac{\Gamma_{An;A'B}
  \Gamma_{nB;AA'}}{E_{n}+E_{A'}-E_{A}-E_{B}}\right]\quad .
\end{equation}

Here $B$, $n$, and $A$ denote initial, intermediate and final 
electron states, respectively. 
$A'$ lables the resonant intermediate state. Then
$\Gamma_{A'}$ is the total width and $\Gamma_{A'A}$,
$\Gamma_{BA'}$ are partial widths, connected with the transitions
$A' \rightarrow A$ and $B \rightarrow A'$. The notations
$\Gamma_{AB;CD}$ are used for the "mixed" transition probabilities
where one amplitude corresponds to the transition $A \rightarrow
C$ and the other one to the transition $B \rightarrow D$. Finally,
$E_{B}$, $E_{A'}$ and $E_{n}$ denote the one-electron energies. In case
of the Lyman-$\alpha$ transition $B=A=1s, A'=2p$ and the photon
frequency is close to $\omega = E_{2p}-E_{1s}$. Then
$\Gamma_{A'}=\Gamma_{A'A}=\Gamma_{BA'}=\Gamma_{2p}$. Assuming that
the "mixed" probabilities are parametrically of the same order as
$\Gamma_{2p} \simeq m \alpha (\alpha Z)^{4}$ 
(in relativistic units) and using
the parametric estimate $\Delta E\simeq m(\alpha Z)^{2}$ for the
energy denominators in Eq. (3) we obtain immediately the estimate
(1). An explicit summation over the total energy spectrum for the
hydrogen atom leads to the coefficient $c_{1}$ quoted above.

Expression (3) corresponds to the interference term between
resonant and nonresonant contributions to the photon scattering
amplitude. The nonresonant contributions arise when we replace the
resonant intermediate state $2p$ by $np$ states with $n>2$ and also when
the incident and emitted photons are interchanged.

In \cite{Jentschura01} it has been observed that a large NR
contribution arises when taking into account the fine structure
of the H atom and considering the state $2p_{3/2}$ as a nonresonant one.
Then the enhancement follows from the small energy denominator
$\Delta E_{f} = E_{2p_{3/2}}-E_{2p_{1/2}}$. However, this
contribution vanishes in the total cross section after angular
integration and remains only finite if the differential cross section is
considered. The latter may correspond to a particular
experimental situation but it is the total cross section which defines
the absolute limit for the accuracy of the frequency measurement.
Additionally, we would like to note 
that there exists also a quadratic NR contribution
to the total cross section from $2p_{3/2}$ state. This
contribution does not vanish after angular integration and is
given by the formula:

\begin{equation}
\delta_{2}=\frac{(\frac{1}{2}\Gamma_{2p})^{4}}{\Delta E_{f}^{3}}\quad .
\end{equation}
This can be easily obtained in the same way as Eq. (3) (see
\cite{LSPS01}). The parametric estimate Eq. (2) immediately follows
from Eq. (4) and from the estimate $\Delta E_{f}=m(\alpha Z)^{4}$.

Formula (3) is valid also for the NR correction to the process
of the resonant two-photon excitation $1s+2\gamma \rightarrow 2s$.
In this case $B=A=1s, A'=2s$, $2\omega =E_{2s}-E_{1s}$,
$\Gamma_{A'} \simeq \Gamma_{A'A} \simeq
\Gamma_{B'A}=\Gamma_{2s,2\gamma}$. Here $\Gamma_{2s,2\gamma}$ is
the two-photon width of the $2s$ level 
(neglecting the small contribution of
the one-photon width $\Gamma_{2s,1\gamma}$). Using the parametric
estimate for the two-photon width $\Gamma_{2s,2\gamma}\simeq m
\alpha^{2}(\alpha Z)^{6}$  one obtains
\begin{equation}
|\delta|=\frac{\Gamma_{2s,2\gamma}^{2}}{\Delta E}\simeq \frac{[m
\alpha^{2}(\alpha Z)^{6}]^{2}}{m (\alpha Z)^{2}}=m
\alpha^{4}(\alpha Z)^{10}
\end{equation}

This estimate has been derived  in Ref. \cite{Jentschura01} where the
conclusion has been drawn, that NR corrections for $1s+2\gamma
\rightarrow 2s$ transition enter at the level $10^{-14}$ Hz and
thus are negligible at the current and projected level of experimental
accuracy. 
At present the latter is $\pm 46$ Hz and is expected
to be two orders of magnitude smaller in future experiments
\cite{Niering00}.

The NR correction (5) corresponds to the process described by the
Feynman graph of Fig. 1. However, the experiment in
\cite{Hunder99,Niering00} is based on a different process. In
this experiment the H atom being in its ground state is excited 
first by the
absorption of the two laser photons into the $2s$ state. 
After some time delay $t_D$ the $2s$ state 
is detected by applying a "small" electric
field and observing the quenched Lyman-$\alpha$ decay. The
corresponding process is described by the Feynman graph of Fig. 2. For
the analysis we will assume that for the "small" electric field
the Stark parameter 
$\xi_{\rm S} = \langle 2s|\vec{d}\cdot\vec{E}|2p\rangle /\Delta E_{\rm L}$ 
corresponds to $\xi_{\rm S}\ll 1$. $\vec{d}$ is the
electric dipole moment operator, $\vec{E}$ is the electric field
strength, $\Delta E_{\rm L}$ is the Lamb shift between $2s$ and $2p$
states. 
Accordingly, we can replace the intermediate $2s$ state in the graph in
Fig. 2 by the state $2s'=2s+\xi_{\rm S}2p$. The resonant cross-section
corresponding to the Feynman graph of Fig. 2 is determined by 

\begin{equation}
\sigma_{{\rm res}} = \frac{1}{2\pi}\,
\frac{\Gamma_{2s,2\gamma}\,\xi_{\rm S}^{2}\,\Gamma_{2p,1\gamma}}{(E_{2s}+\Delta
E_{s}^{(2)}-E_{1s}-2\omega)^{2}+
\frac{1}{4}\xi_{\rm S}^{4}\,\Gamma_{2p,1\gamma}^{2}}\quad .
\end{equation}

Here $\Gamma_{2p,1\gamma}$ denotes the Lyman-$\alpha$ width and $\Delta
E_{s}^{(2)}$ is the second-order Stark shift for the $2s$ state. The
Stark shift for the $1s$ state can be neglected for "small" electric
fields. However, in the real experiment \cite{Hunder99,Niering00}
the excitation region is separated spatially from the detection
region. Therefore the Stark shift does not contribute to the excitation
condition: $2\omega =E_{2s}-E_{1s}$. Moreover, the width of the
resonance is defined by the time delay $t_{D}$, which is necessary
for the atom to reach the detection region: for $\xi_{\rm S}\sim 0.1$
the decay time in the electric field is
$(\xi_{\rm S}^{2}\Gamma_{2p,1\gamma})^{-1}\simeq 10^{-7}$ s. This is very
small time compared to $t_{D}$: the characteristic atomic
velocities are $10^{4}$ cm/s and the space separation of the
excitation and detection regions is $13$ cm
\cite{Hunder99,Niering00}. Then $t_{D}\simeq 10^{-3}$ s what
corresponds to the experimental width of the resonance
$\Gamma_{{\rm exp}}\sim 1$ kHz. Then Eq. (6) should be replaced by

\begin{equation}
\sigma_{{\rm res}}=\frac{1}{2\pi}\,
\frac{\Gamma_{2s,2\gamma}\,\Gamma_{{\rm exp}}}{(E_{2s}-
E_{1s}-2\omega)^{2}+\frac{1}{4}\Gamma_{{\rm exp}}^{2}}\quad .
\end{equation}

The major nonresonant contribution arises from the closely lying
$2p$ level

\begin{equation}
\sigma_{{\rm NR}}=\frac{1}{2\pi}\,
\frac{\Gamma_{2p,2\gamma}\,\Gamma_{2p,1\gamma}}{(E_{2p}
-E_{1s}-2\omega)^{2}+\frac{1}{4}\Gamma_{2p,1\gamma}^{2}}
\end{equation}
where $\Gamma_{2p,2\gamma}$ is the two-photon width of the $2p$
level. The interference term between $\sigma_{{\rm res}}$ and
$\sigma_{{\rm NR}}$ is absent in the total cross section, but may be
present in the differential cross section.

To find the NR correction we employ the same idea as in \cite{LSPS01},
i.e. we determine the position of the maximum for the function $\sigma
(\omega)$.

In the case in which the total cross section is measured we obtain

\begin{equation}
\delta=  \frac{\Gamma_{2p,2\gamma} 
\Gamma_{2p,1\gamma}}{\Gamma_{2s,2\gamma}\Gamma_{{\rm exp}}}
\,\frac{(\frac{1}{2}\Gamma_{{\rm exp}})^{4}}{\Delta E_{\rm L}^{3}}
\end{equation}

Here we assumed that $\Delta E_{\rm L}\gg \Gamma_{2p,1\gamma}$ where
$\Delta E_{\rm L}=E_{2s}-E_{2p}$ is the Lamb shift. The corresponding
expressions (in relativistic units) 
are: $\Gamma_{2p, 1\gamma}=\left(
\frac{2}{3}\right)^{8}m\alpha (\alpha Z)^{4}=0.04 m\alpha (\alpha
Z)^{4}$ and $\Delta E_{\rm L}=0.4m\alpha (\alpha Z)^{4}$. Taking into
account that $\Gamma_{2p,2\gamma}$ is defined by the E1M1 transition
and $\Gamma_{2s,2\gamma}$ corresponds to the 2E1 transition, we have
$\Gamma_{2p,2\gamma}/\Gamma_{2s,2\gamma}\simeq (\alpha Z)^{2}$.
Insertion of these quantities into Eq. (9) leads to the negligible NR
correction. The NR correction to the differential cross-section is
much larger. In this case the formula of the type of Eq. (3) for
differential cross section should be used:

\begin{equation}
|\delta|=\frac{1}{2}\left( \frac{\Gamma_{2p,2\gamma} \Gamma_{2p,
1\gamma}}{\Gamma_{2s,2\gamma}\Gamma_{{\rm exp}}}\right)^{\frac{1}{2}}
\frac{(\Gamma_{{\rm exp}})^{2}}{\Delta E_{\rm L}}\simeq 10^{-3}\, {\rm Hz} \quad .
\end{equation}

This limit for the experimental accuracy turns out to be many orders of
magnitude larger than the corresponding value  obtained from (5). 
It is the value (10) which should be compared with
the projected experimental accuracy of about $10^{-1}$ Hz
\cite{Hunder99,Niering00}.
\begin{center}
Acknowledgements
\end{center}

The work of L.L. and D.S. was supported by the RFBR Grant $No.$
99-02-18526 and by the Minobrazovanie grant $No.$ E00-3.1.-7.
G. P. and G. S. acknowledge financial support from BMBF, DFG and GSI.

\newpage
\begin{figure}
\centerline{\mbox{\epsfxsize=7cm \epsffile{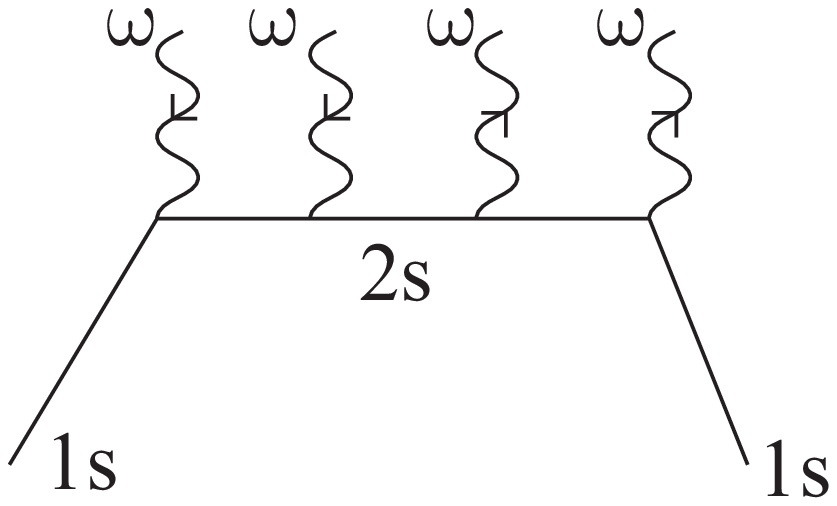}}}
\caption{
The Feynman graph for the resonance excitation of $2s$ level in the
double photon scattering process $1s+2\gamma \rightarrow 2s$. The
solid line denotes the electrons, the wavy lines with the arrows
denote the incident and emitted photons.}
\end{figure}

\begin{figure}
\centerline{\mbox{\epsfxsize=7cm \epsffile{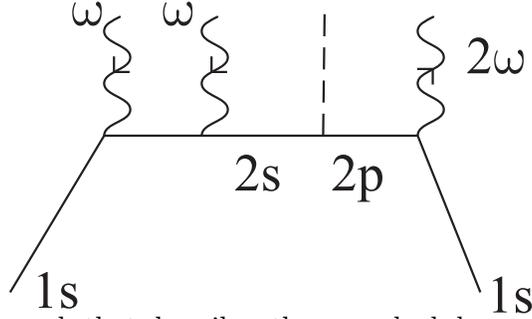}}}
\caption{
The Feynman graph that describes the quenched decay of the
intermediate $2s$ state. The electric field is denoted by the dashed
line. The other notations are the same as in Fig. 1.}
\end{figure}

\end{document}